\documentclass[twocolumn,showpacs,preprintnumbers,amsmath,amssymb]{revtex4-1}
\usepackage{graphicx}
\usepackage{dcolumn}
\usepackage{mathrsfs}
\usepackage{bm}
\usepackage{lineno}

\begin{document}
\bibliographystyle{osajnl}


\title{Joint Measurement of Multi-Parameter}

\author{Jiamin Li$^{1}$}
\author{Yuhong Liu$^{1}$}
\author{Liang Cui$^{1}$}
\author{Nan Huo$^{1}$}
\author{Syed M Assad$^{3}$}
\author{Xiaoying Li$^{1}$}
 \email{xiaoyingli@tju.edu.cn}
\author{Z. Y. Ou$^{1, 2,}$}
 \email{zou@iupui.edu}
\affiliation{%
$^{1}$College of Precision Instrument and Opto-Electronics Engineering, Key Laboratory of
Opto-Electronics Information Technology, Ministry of Education, Tianjin University,
Tianjin 300072, P. R. China\\
$^{2}$Department of Physics, Indiana University-Purdue University Indianapolis, Indianapolis, IN 46202, USA\\
$^{3}$Department of Quantum Science, The Australian National University, Canberra ACT 0200, Australia
}%

\date{\today}

\begin{abstract}
Although quantum metrology allows us to make precision measurement beyond the standard quantum limit, it mostly works on the measurement of only one observable due to
Heisenberg uncertainty relation on the measurement precision of non-commuting observables for one system. In this paper, we study the schemes of joint measurement of multiple observables which do not commute with each other by using the quantum entanglement between two systems. We focus on analyzing the performance of newly developed SU(1,1) interferometer on fulfilling the task of joint measurement. The results show that the information encoded in multiple non-commuting observables on an optical field can be simultaneously measured with a signal-to-noise ratio higher than the standard quantum limit, and the ultimate limit of each observable is still the Heisenberg limit. Moreover, we find a resource conservation rule for the joint measurement.

\end{abstract}

\maketitle

\section{introduction}

Quantum metrology, which uses quantum resources to improve the sensitivity beyond the classical limit in the estimation of relevant physical parameters, has been one of the frontier topics in the applications of quantum technology \cite{cav,fri85}.
Most of the previous studies are committed to improve the signal-to-noise ratio (SNR) of a single parameter, such as precision measurement of a phase shift, and the basic idea is to reduce the quantum noise in the measurement with novel quantum states \cite{xiao,gran}. So far, squeezed state has been widely applied in quantum precision measurement of single parameter, such as gravitational wave detection \cite{ligo}. In some applications, however, the information is embedded in two or more non-commuting observables. For example, information about the real and imaginary parts of the linear susceptibility of an optical medium is embedded in the phase and amplitude of a probe optical field passing through the medium in the form of small modulated signals. According to the Heisenberg uncertainty principle on two non-commuting observables, quantum noise reduction in one observable is inevitably accompanied by the noise increase in the other. Therefore, the strategy of quantum noise reduction fails in measuring two non-commuting observables with sensitivity simultaneously  higher than classical limit.

On the other hand, Heisenberg uncertainty involves in two conjugate quantities of one system, but the situation for two systems is completely different. Quantum entanglement allows perfect quantum correlations between two systems. Einstein, Podolsky, and Rosen (EPR) showed in a seminal paper \cite{epr} that there exists such a state of two particle systems that exhibits perfect correlations not only between the positions of two remotely located particles but also between their momenta. This is so because the difference of their position operators $\hat x_1 - \hat x_2$ and the sum of their momenta operators $\hat p_1 + \hat p_2$ commute: $[\hat x_1 - \hat x_2, \hat p_1 + \hat p_2]=0$. Such perfect correlations led to the famous EPR paradox about the incompleteness of quantum mechanics via a locality argument. The experimental realization of the EPR entangled state and the demonstration of EPR paradox were first done in an optical system of non-degenerate parametric amplifier \cite{reid,ou92} in which the two particles are the virtual harmonic oscillators representing two spatially separated modes of optical beams with $\hat x_{1,2} \propto \hat a_{1,2}^{\dag} + \hat a_{1,2} \equiv \hat X_{1,2}$ and $\hat p_{1,2} \propto i(\hat a_{1,2}^{\dag} - \hat a_{1,2})\equiv \hat Y_{1,2}$, where $\hat a_{i}^{\dag}$ and $\hat a_{i}$ ($i=1,2$) are the creation and annihilation operators of the two optical fields. These magic quantum nonlocal correlations of orthogonal observables give rise to quantum noise reduction by noise cancelation via $\hat X_1 - \hat X_2$ and $\hat Y_1+\hat Y_2$ and can be employed for the simultaneous measurement of the phase and amplitude encoded in $\hat Y_1$ and $\hat X_1$ of one optical beam. This idea was first proposed in the form of quantum dense coding \cite{brau,zh} and was demonstrated experimentally in the joint measurement of two orthogonal observables with precision beating standard quantum limit (SQL) \cite{xyl,sna}.

Along a similar line of argument, the quantum entanglement has also been used in quantum noise cancelation in an amplifier for noiseless quantum amplification \cite{ou93,kong13}. This is exactly the underlying principle for the so-called SU(1,1) interferometer (SUI), a new type of nonlinear interferometer that is based on nonlinear parametric processes for wave splitting and superposition. Proposed as early as in 1986 by Yurke et al. \cite{yur}, the SU(1,1) interferometer can in principle reach the Heisenberg limit in the precision measurement of phase shift \cite{ou97,PDL17b}. 
Although practical imperfections limited its ability to reach the ultimate precision, it was demonstrated that SU(1,1) interferometers can still beat the standard quantum limit of phase measurement and are superior to traditional interferometers in a number of ways \cite{hud14,PDL12,PDL17a,PDL17b,pl}.

So far, for the SU(1,1) interferometer used for the quantum enhanced phase measurement, only one of the two output ports is exploited. However, it turns out that the other output port of SUI also contains the information of the sensing field inside the interferometer \cite{Guo2016}, which can be used for amplitude measurement. 
In this paper, we will study the performance of SU(1,1) interferometer in the application of simultaneously measuring non-commuting observables  with precision beating SQL. We will find the optimum operation condition  for achieving the highest signal-to-noise ratio in the simultaneous measurement of each observable. Also, we will compare this scheme of joint measurement with others using classical light. Moreover, we will compare the performance of SU(1,1) interferometer respectively with one-beam and dual-beam function as the sensing field, which leads to a resource conservation rule for joint measurement. Furthermore, we will demonstrate that the ultimate limit of the precision in the joint measurement is still the Heisenberg limit.

The rest of the paper is organized as follows. We first briefly review the signal-to-noise ratio of multi-parameter measurement obtained by using three typical  classical schemes in Sec. II. 
Next, in Sec. III, we study the quantum enhanced measurement schemes by using quantum entanglement, including quantum dense coding scheme and SU(1,1) nonlinear interferometer. In this section, we focus on analyzing the SU(1,1) interferometer as a platform for joint measurement and demonstrating its advantages over the dense coding scheme. 
In Sec. IV, we discuss how the Heisenberg limit can be approached in both phase and amplitude measurement. Finally, we briefly conclude in Sec. V.

\section{Joint measurement schemes with classical light}

Before introducing the quantum enhanced joint measurement schemes, we start by first considering the measurement schemes using classical light sources.
These give rise to SQL for the joint measurement of the information embedded in two non-commuting quadrature-phase amplitudes of an optical beam.

\subsection{Direct measurement}
\begin{figure}[htb]
\centering
 \includegraphics[width=7cm]{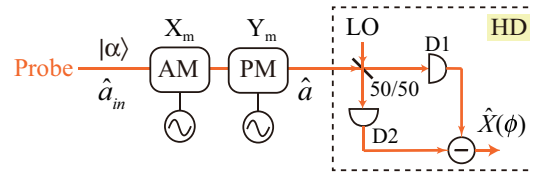}
\caption{Direct Measurement Scheme. AM, amplitude modulator; PM, phase modulator; HD, homodyne detection; D1-D2, detectors; LO, local oscillator. }\label{fig-DM}
\end{figure}

To obtain the information of phase or amplitude carried by a probe beam, the simplest scheme is the direct measurement realized by using homodyne detection, as shown in Fig. \ref{fig-DM}. 
We assume the probe field $\hat a_{in}$ is in a coherent state $|\alpha\rangle$ with $|\alpha \rangle= |\alpha|e^{j\varphi_0}$ where $\alpha$ is a complex number and $\varphi_0$ is the initial phase. 
By passing the probe through an amplitude modulator (AM) and a phase modulator (PM), a weak phase modulation of $\delta \ll 1$ and a weak amplitude modulation $\epsilon \ll 1$ simultaneously applied to the probe beam can be expressed as $e^{j\delta} \approx 1 + j\delta$ and $e^{-\epsilon} \approx  1 - \epsilon$, respectively. 
The modulated probe field is then expressed as $\hat a = \hat a_{in}e^{j\delta}e^{-\epsilon} \approx \hat a_{in}(1+j\delta-\epsilon)$. 
On the other hand, a modulation signal of an arbitrary quadrature amplitude $X_m(\theta) = X_m\cos\theta + Y_m\sin\theta$ can be viewed as a combination of amplitude and phase modulations, where $X_m = X_m(0)\equiv \epsilon$ and $Y_m = X_m(\pi/2)\equiv \delta$. 
Here, for the consistency with the analysis here-in-after, $X_m$ and $Y_m$ are used to respectively denote the amplitude modulation and phase modulation, while $X_m(\theta)$ is used to specify a modulation signal of quadrature amplitude at arbitrary angle $\theta$. 
The modulated signal of probe beam can be directly measured by using homodyne detection (HD), which consists of a 50/50 beam splitter and two detectors (D1 and D2). When the phase of local oscillator (LO) is set to $\phi$, the difference between the photocurrents of D1 and D2 gives the measurement of the modulated probe beam $\hat X(\phi) = \hat a e^{-j\phi}+\hat a^{\dag}e^{j\phi}$.

Defining $\hat X \equiv \hat X(\varphi_0)$ and $\hat Y \equiv \hat X(\varphi_0+\pi/2)$ as two conjugate observables, it is straightforward to deduce that the measurement on probe beam for $\hat X$ gives the amplitude modulation signal $\langle \hat X\rangle = 2|\alpha|\epsilon =2|\alpha|X_m$, whereas that for $\hat Y$ gives the phase modulation $\langle \hat Y\rangle = 2|\alpha|\delta = 2|\alpha|Y_m$. In general, the measurement of $\hat X(\varphi_0+\theta)$ measures the modulated quadrature amplitude at arbitrary angle  $\langle \hat X(\varphi_0+\theta)\rangle = 2|\alpha|X_m(\theta)$. On the other hand, for the probe in coherent state, its noise is independent upon the phase of LO, i.e., $\langle \Delta^2\hat X(\phi)\rangle =1$. So we have the SNR for the direct measurement of phase or amplitude:
\begin{equation}
\begin{split}
SNR_{DM}(\hat Y) = \frac{\langle \hat Y \rangle^2}{\langle \Delta \hat Y^2 \rangle}=4I_{ps}\delta^2 ~~{\rm or} \\ SNR_{DM}(\hat X) =\frac{\langle \hat X \rangle^2}{\langle \Delta \hat X^2 \rangle} =4I_{ps}\epsilon^2 , \label{SNR-XY0}
\end{split}
\end{equation}
where $I_{ps} \equiv |\alpha|^2$ is the intensity or photon number for the probe sensing beam and the subscript ``DM'' represents the direct measurement scheme in Fig. \ref{fig-DM}.

\subsection{Beam splitting method}

\begin{figure}[htb]
\centering
 \includegraphics[width=7cm]{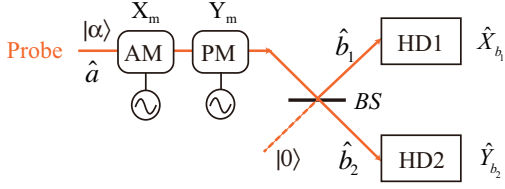}
\caption{Joint measurement scheme using beam splitter (BS). AM, amplitude modulator; PM, phase modulator; HD, homodyne detection.}\label{fig-BS}
\end{figure}

However, the scheme in Fig. \ref{fig-DM} can only make direct measurement on one observable at one time, i.e., either phase or amplitude can be measured at one time.
A straightforward method of jointly measuring $\hat X$ and $\hat Y$ is to split the modulated probe beam into two with a beam splitter (BS), as shown in Fig. \ref{fig-BS}. The two quadrature-phase amplitudes $\hat X$ and $\hat Y$ can be simultaneously measured at the two outputs of BS by using HD1 and HD2, respectively.
Since the probe is in an ideal coherent state $|\alpha\rangle$, whose noise is simply given by $\langle \Delta^2 \hat X(\phi)\rangle =1$ (the same as the vacuum noise). If the detection efficiency of each HD device is perfect, and the relative phase of each HD is properly locked,  the SNRs of simultaneously measured phase and amplitude signals are expressed as 
\begin{equation}
SNR_{BS}(\hat Y_{b_2})=4T I_{ps}\delta^2,~~ SNR_{BS}(\hat X_{b_1})=4R I_{ps}\epsilon^2 \label{SNR-XYTR}.
\end{equation}
where the subscripts ``$b_1$'' and ``$b_2$'' denote the fields at two ports of BS, and $T$ and $R$ with $T+R=1$ are the transmissivity and reflectivity of BS. When the sizes of phase modulation and amplitude modulation are equal, i.e., $\delta=\epsilon$, we have
\begin{eqnarray}
SNR_{BS}(\hat Y_{b_2})+ SNR_{BS}(\hat X_{b_1})&=& SNR_{DM}(\hat Y)\cr &=& SNR_{DM}(\hat X)\label{SNR-X+Y}.
\end{eqnarray}
From Eq. (\ref{SNR-X+Y}), we find the total sum of the SNRs for joint measurement of phase and amplitude is equal to the SNR of phase or amplitude obtained by direct measurement in Fig. \ref{fig-DM}, in which all the resource is consumed on one observable. In other word, the joint measurement can be viewed as the partition of the total resource into two observables.

For the BS with the splitting ratio $T=R=1/2$, we have
\begin{equation}
SNR_{BS}(\hat Y_{b_2})=2I_{ps}\delta^2,~~ SNR_{BS}(\hat X_{b_1})=2I_{ps}\epsilon^2.
\label{SNR-XY}
\end{equation}
Notice that $SNR_{BS}(\hat Y_{b_2}) = SNR_{DM}(\hat Y)/2$ and $SNR_{BS}(\hat X_{b_1}) = SNR_{DM}(\hat X)/{2}$ are referred to as the SQL of the joint measurement. Comparing with the SNR obtained by direct measurement method (see Eq. (\ref{SNR-XY0})), there is a 3dB-reduction for SNR measured at each output port in Fig. \ref{fig-BS}, which is originated from the vacuum $|0 \rangle$ entering from the unused port of the BS.

\subsection{Beam splitting scheme with an optical parametric amplifier}

\begin{figure}[htb]
\centering
 \includegraphics[width=8cm]{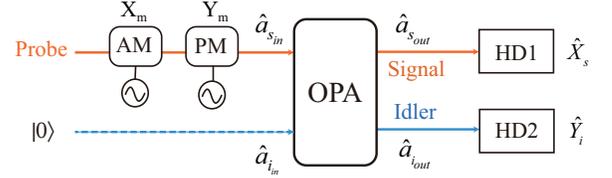}
\caption{Joint measurement scheme using an optical parametric amplifier (OPA) as a beam splitter. AM, amplitude modulator; PM, phase modulator; HD, homodyne detection.}\label{fig-Amp}
\end{figure}

In a real experiment, the measured SNRs are usually smaller than those given by Eq. (\ref{SNR-XY}) due to the non-ideal detection efficiencies of HDs. One way to mitigate the influence of detection loss is to replace the 50/50 BS in Fig. \ref{fig-BS} with a conventional optical parametric amplifier (OPA), as shown in Fig. \ref{fig-Amp}. As we will show in the following, the OPA scheme at high gain is similar to the BS scheme in Fig. \ref{fig-BS}. So SNRs of joint measurement performed by using OPA can be viewed as a direct comparison with the quantum scheme of SU(1,1) interferometer (see Sec. IIIB for detail).

An optical parametric amplifier has its input probe beam amplified at the signal port and in the meantime also outputs another field called idler, which contains the information of the input probe field as well \cite{cav82}. From the input-output relation of the OPA:
\begin{equation}
{\hat a}_{s_{out}} = {G}{\hat a}_{s_{in}} + {g}{{\hat a}_{i_{in}}^\dag },~~ {{\hat a}_{i_{out}}} = {G}{\hat a}_{i_{in}}+ {g}{{\hat a}_{s_{in}}^\dag }.
\label{cd-out}
\end{equation}
with $G^2-g^2=1$, where $G$, $g$ are the amplitude gains of OPA, the signal probe input $\hat a_{s_{in}}$ is in a coherent state and idler input $\hat a_{i_{in}}$ is in vacuum.
The signal and idler output fields have the averages of $\langle {\hat a}_{s_{out}}\rangle = G\alpha$ and $\langle {\hat a}_{i_{out}}\rangle = g\alpha^*$. Thus, an OPA can act as a beam splitter through which the information of phase and amplitude modulation encoded on probe beam is distributed to the signal and idler output ports.

From Eq. (\ref{cd-out}), it is straightforward to deduce the average powers of the phase and amplitude modulations measured by HD1 and HD2,
\begin{equation}
\langle \hat X_{s_{out}}\rangle^2 = 4 G^2I_{ps}\epsilon^2, ~~ \langle \hat Y_{i_{out}}\rangle^2 = 4 g^2I_{ps}\delta^2\label{sig-amp}.
\end{equation}
where the subscripts $s_{out}$ and $i_{out}$ respectively indicate the signal and idler outputs.
Moreover, according to the noise measured at the signal and idler outputs,
\begin{equation}
\langle \Delta^2\hat X_{s_{out}} \rangle  = \langle \Delta^2\hat Y_{i_{out}}  \rangle = G^2+g^2,
\label{noise-amp}
\end{equation}
we arrive at the SNRs of phase and amplitude:
\begin{equation}
SNR_{Amp}({\hat X_s}) = {4G^2I_{ps}\epsilon^2\over G^2+g^2},
~~ SNR_{Amp}({\hat Y_i}) = {4g^2I_{ps}\delta^2\over G^2+g^2}.
\label{SNR-amp}
\end{equation}
where the subscript ``Amp" denotes that the results are for conventional amplifier scheme in Fig. \ref{fig-Amp}. In the case of $G \to \infty$, SNRs in Eq. (\ref{SNR-amp}) are rewritten as
\begin{equation}
\begin{split}
SNR_{Amp}({\hat X_s}) = {4G^2I_{ps}\epsilon^2\over G^2+g^2} \rightarrow 2I_{ps}\epsilon^2\\
SNR_{Amp}({\hat Y_i}) = {4g^2I_{ps}\delta^2\over G^2+g^2} \rightarrow 2I_{ps}\delta^2.
\label{SNRopa}
\end{split}
\end{equation}
It is obvious that the SNRs in Eq. (\ref{SNRopa}) are the same as the SQL in Eq. (\ref{SNR-XY}).
Notice that for the case of $\delta=\epsilon$, similar to Eq. (\ref{SNR-X+Y}), we  again have the resource partition relation:
\begin{equation}
SNR_{Amp}({\hat X_s}) + SNR_{Amp}({\hat Y_i}) = SNR_{DM}({\hat X}).
\label{SNR-res}
\end{equation}

\begin{figure}[htb]
\centering
 \includegraphics[width=6.5cm]{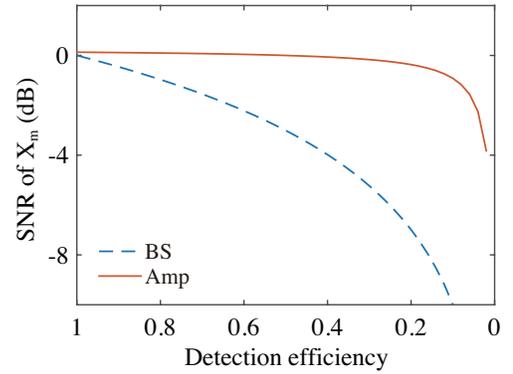}
\caption{Signal-to-noise ratio (SNR) of the measured amplitude $X_m$ as a function of detection efficiency $1-L_d$. Solid curve and dashed curve respectively represent the joint measurement results for the schemes in Fig. \ref{fig-Amp} and Fig. \ref{fig-BS}. In the calculation, $I_{ps}\epsilon^2 = 1/2$, and the amplitude gain of OPA is $g=5$.}
\label{loss-amp}
\end{figure}

To illustrate the loss tolerance advantage of OPA scheme, let's analyze the influence of detection efficiency on the SNRs of joint measurement. 
In general, the detection loss $L_d$ is modeled by placing a beam splitter in front of each HD. 
The transmissivity of the beam splitter is viewed as $1-L_d$, and vacuum field $\hat v$ is coupled into the detected field through the non-ideal transmissivity. 
Taking the signal output of OPA as an example, the operator of the detected field is given by: $\hat a'_{s_{out}} = \sqrt{(1-L_d)} \hat a_{s_{out}} + \sqrt{L_d} \hat v$.
So the measured average power of amplitude,
\begin{equation}
\langle \hat X'_{s_{out}} \rangle ^2 = (1-L_d)\langle \hat X_{s_{out}} \rangle^2,
\label{Sigloss}
\end{equation}
decreases with the increase of detection loss. Meanwhile, the measured noise
\begin{equation}
\langle \Delta^2 \hat X'_{s_{out}} \rangle  = (1-L_d)\langle  \Delta^2 \hat X_{s_{out}} \rangle + L_d
\label{noiseloss}
\end{equation}
accordingly changes because of the vacuum noise coupled in through loss.
Since the noise of OPA (see Eq. (\ref{noise-amp})) is much larger than that of vacuum, particularly in the high gain regime, the noise in Eq. (\ref{noiseloss}) decreases with the increase of $L_d$, and the decrease rate is about the same as that of the signal power in Eq. (\ref{Sigloss}). Therefore, the SNR of the joint measurement scheme in Fig. \ref{fig-Amp} is not sensitive to detection loss when the amplitude gain of OPA and the detection efficiency of HD are not too low, as shown by the solid curve in Fig. \ref{loss-amp}. As a comparison, we also plot SNR of amplitude modulation measured by BS scheme (Fig. \ref{fig-BS}) as a function of the detection efficiency, as shown by the dashed curve in Fig. \ref{loss-amp}. Clearly, the SNRs for the BS scheme significantly decrease with the detection efficiency. This is because the noise at each output of BS scheme is always at the vacuum noise level, while the measured average power of signal modulation is proportional to the detection efficiency.

\section{Joint measurement schemes with quantum fields}

In this section, we will analyze the quantum enhanced joint measurement by briefly reviewing the quantum dense coding scheme for quantum dense metrology at first. 
Then we will focus on analyzing the SU(1,1) interferometer (SUI) in the application of joint measurement.
In addition to studying its optimum operation condition for achieving the maximized SNRs in joint measurement, we will compare the performance of SUI with one-beam and dual-beam function as the sensing field. Finally, we will discuss the resource conservation rule when two non-commuting observables are simultaneously measured.

\subsection{Quantum dense coding scheme for quantum dense metrology}

\begin{figure}[htb]
\centering
 \includegraphics[width=8.5cm]{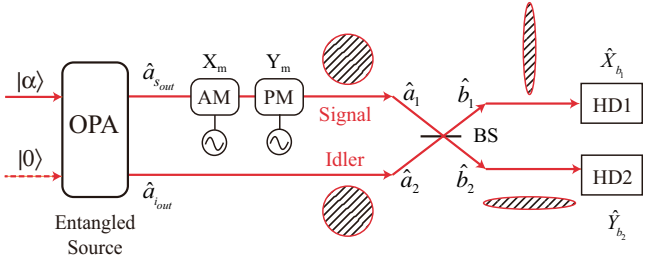}
\caption{Dense coding scheme for quantum dense metrology. OPA, optical parametric amplifier; AM, amplitude modulator; PM, phase modulator; HD, homodyne detection. The shadows illustrate the noise distribution of the fields at the two inputs and outputs of BS. }\label{fig-Qcode}
\end{figure}

To implement the quantum enhanced joint measurement, we need to design a simultaneous measurement on both $\hat X_1 - \hat X_2$ and $\hat Y_1+\hat Y_2$ of a pair of entangled fields to extract the information carried by probe. 
This is the quantum dense coding scheme \cite{brau} shown in Fig. \ref{fig-Qcode}. The two EPR entangled fields labeled as $\hat a_{s_{out}},\hat a_{i_{out}}$ are generated from an OPA \cite{reid,ou92,xyl}. 
One of the entangled fields are encoded with both phase and amplitude information by a phase modulator (PM) and an amplitude modulator (AM), respectively. 
When the modulated field and the other half of entangled field, now labeled as $\hat a_1, \hat a_2$,  are superimposed at a 50/50 beam splitter, the two outputs of the BS are given by
\begin{equation}
\hat b_1= (\hat a_1- \hat a_2)/\sqrt{2}, ~~\hat b_2= (\hat a_2+ \hat a_1)/\sqrt{2}.
\end{equation}
If we measure $\hat X_{b_1}= (\hat X_{a_1}-\hat X_{a_2})/\sqrt{2}$ at $\hat b_1$ port and $\hat Y_{b_2} =(\hat Y_{a_1}+\hat Y_{a_2})/\sqrt{2}$ at $\hat b_2$ port by using HD1 and HD2, respectively, we can achieve the simultaneous measurement of $\hat X_{a_1}-\hat X_{a_2}$ and $\hat Y_{a_1}+\hat Y_{a_2}$ \cite{ou92,brau,zh,xyl}. So the amplitude and phase modulation signals carried by $\hat a_1$, can be obtained by simultaneously measuring $\hat X_{b_1}$ and $\hat Y_{b_2}$, whose noise fluctuations are lower than shot noise level (SNL), and the measurement sensitivities are beyond SQL. 

In Fig. \ref{fig-Qcode}, the two outputs of the OPA are described by
\begin{equation}
{\hat a}_{s_{out}} = {G}{\hat a}_{s_{in}} + {g}{{\hat a}_{i_{in}}^\dag },~~ {{\hat a}_{i_{out}}} = {G}{\hat a}_{i_{in}} + {g}{{\hat a}_{s_{in}}^\dag },
\label{AB-out}
\end{equation}
where $\hat a_{s_{in}}$ in coherent state $|\alpha\rangle$ is the weak signal input, and $\hat a_{i_{in}}$ in vacuum state $| 0 \rangle$ is the idler input.
The modulated signal and its correlated field before BS, expressed as $\hat a_1=\hat a_{s_{out}}(1+j\delta-\epsilon)$ and $\hat a_2=\hat a_{i_{out}}$, are then combined by BS.
At the two outputs of BS, the operators  $\hat X_{b_1} = \hat b_1e^{-j\varphi_0} + \hat b_1^{\dag}e^{j\varphi_0}$ and $\hat Y_{b_2} = (\hat b_2e^{-j\varphi_0} - \hat b_2^{\dag}e^{j\varphi_0})/j$ ($e^{j\varphi_0}\equiv \alpha/|\alpha|$) are measured by HD1 and HD2 with $\phi_1=0$ and $\phi_2=\pi/2$, respectively. The measurement gives the powers of modulated amplitude and phase signals:
\begin{equation}
\langle \hat X_{b_1}\rangle^2 = 2 I_{ps}\epsilon^2,~~ \langle \hat Y_{b_2}\rangle^2 = 2I_{ps}\delta^2,
\end{equation}
where ${I_{ps}} = \left\langle {\hat a_{s_{out}}^{\dag}\hat a_{s_{out}}} \right\rangle = G_1^2{\left| \alpha  \right|^2} ~({\left| \alpha  \right|^2} \gg 1)$, with $\left| \alpha  \right|^2$ denoting the intensity of the seed injection, is the intensity of the probe beam.
Meanwhile, the noise fluctuations measured at the two outputs of BS are given by
\begin{equation}
\begin{split}
\langle \Delta^2\hat X_{b_1}\rangle = \langle \Delta^2(\hat X_{a_1}-\hat X_{a_2})\rangle = 1/{(G+g)^2},\\
\langle \Delta^2\hat Y_{b_2}\rangle = \langle \Delta^2(\hat Y_{a_1}+\hat Y_{a_2})\rangle = 1/{(G+g)^2}, 
\label{noise-EPR11}
\end{split}
\end{equation}
which are lower than the SNL due to the entanglement correlation between the fields $\hat a_1$ and $\hat a_2$. Accordingly, the SNRs of the jointly measured amplitude and phase modulations are
\begin{equation}
\begin{split}
SNR_{DC}(\hat X_{b_1}) = {2 (G+g)^2I_{ps}\epsilon^2},\\
SNR_{DC}(\hat Y_{b_2})= {2 (G+g)^2 I_{ps}\delta^2 },\label{SNR-EPR1}
\end{split}
\end{equation}
where the subscript ``DC" is used to denote dense coding scheme. Comparing with the SQLs in Eq. (\ref{SNR-XY}), it is obvious that the SNRs of joint measurement are improved by using the EPR entanglement with a factor of $(G+g)^2$. Moreover, it is worth noting that for the dense coding scheme, it is critical to carefully set the LOs of HD1 and HD2 to achieve the best quantum enhancement, because noise at each output of BS highly depends on the phase of LOs, $\phi_1$ and $\phi_2$, as shown by the shadows in Fig. \ref{fig-Qcode}.

\subsection{Non-degenerate SU(1,1) Interferometer with one beam sensing}

\begin{figure*}[htb]
\centering
 \includegraphics[width=15cm]{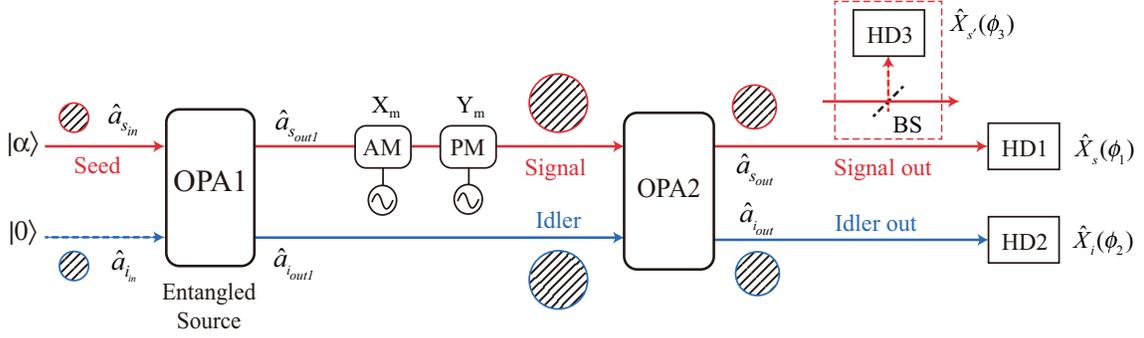}
\caption{
SU(1,1) interferometer formed by OPA1 and OPA2 for the joint measurement of multiple non-commuting quadratures.
 AM, amplitude modulator; PM, phase modulator; OPA, optical parametric amplifier; HD, homodyne detection; BS, beam splitter. The shadows illustrate the noise distribution at different ports of OPAs. }\label{fig-SUI}
\end{figure*}

The BS in dense coding scheme in Fig. \ref{fig-Qcode} can only realize the coherent combination of two entanglement fields with same frequency. To  coherently combine two entangled fields with non-degenerate frequencies, we resort to an SU(1,1) interferometer (SUI), in which the BS in Fig. \ref{fig-Qcode} is replaced with an optical parametric amplifier.

As shown in Fig. \ref{fig-SUI}, the SU(1,1) interferometer consists of two OPAs, which respectively act as beam splitters for wave splitting and superposition. What makes it different from a conventional linear interferometer is that the two fields splitted by OPA1 are now correlated in noise, which can be canceled out in the second OPA due to destructive quantum interference. When the modulation signals encoded in the signal beam out of OPA1 is amplified by OPA2, a noiseless amplification can be achieved at each output of SUI \cite{Guo2016}.


In Fig. \ref{fig-SUI}, the seed injection and vacuum respectively at two input ports of OPA1 are denoted as the field operators $\hat a_{s_{in}}$ and $\hat a_{i_{in}}$. The two entangled quantum fields out of OPA1, are referred to as $\hat a_{s_{out1}}$ and $\hat a_{i_{out1}}$, respectively. The signal probe field, encoded with the information of multiple non-commuting observables by successively passing through an AM and a PM, is sent into the OPA2 together the idler field. The information carried by the probe beam is then amplified by OPA2, whose outputs are denoted as $\hat a_{s_{out}}$ and $\hat a_{i_{out}}$, respectively.


The theoretical analyses in Refs. \cite{Guo2016} and \cite{ou12} show that OPA2 of SUI functions as a phase insensitive amplifier for the information carried by the signal probe beam. When the probe is embedded with the weakly modulated phase and amplitude signals, $\delta \ll 1$ and $\epsilon \ll 1$, the average powers of the two observables respectively measured at the signal and idler output ports by HD1 and HD2 are given by:
\begin{equation}
\langle \hat X_{s_{out}}\rangle^2  = 4G_2^2{I_{ps}}\epsilon^2, ~~\langle \hat Y_{i_{out}}\rangle^2  =4g_2^2{I_{ps}}\delta^2,
\label{sig}
\end{equation}
where ${I_{ps}} = \left\langle {\hat a_{s_{out1}}^{\dag}\hat a_{s_{out1}}} \right\rangle = G_1^2{\left| \alpha  \right|^2} ~({\left| \alpha  \right|^2} \gg 1)$. However, the noise fluctuation at each output port of SUI is sensitive to the relative phase between the pump and the two input fields of OPA2. For brevity, the relative phase is represented by introducing a phase shift $\varphi$ to the idler field $\hat a_{i_{out1}}$.
When OPA2 is operated at the deamplification condition, i.e., $\varphi=\pi$ \cite{ou12,Guo2016}, the intensities at the two outputs of SUI are minimum. In this case, the noise fluctuation measured at each output port takes the minimum and is expressed as
\begin{eqnarray}
\langle {{\Delta ^2}{{\hat  X}_{s_{out}}(\phi_1)}} \rangle
&= & {\left( {{G_2}{G_1} - {g_1}{g_2}} \right)^2} + {({G_1}{g_2} - {G_2}{g_1})^2}\cr
& = & \langle {{\Delta ^2}{{\hat  X}_{i_{out}}(\phi_2)}} \rangle ,
\label{noise}
\end{eqnarray}
where $g_1$ and $g_2$, satisfying the relation $G_{k}^2 - g_{k}^2=1$ ($k=1,2$), are the amplitude gains of OPA1 and OPA2, respectively. Notice that the output noise of SUI is independent of the quadrature phase angles $\phi_1,\phi_2$ of LOs in homodyne detections, as represented by the circularly shaped shadows in Fig. \ref{fig-SUI}. This is a unique property of SUI.

In the case of $g_1=g_2$, the noise fluctuation takes the absolute minimum $\langle {{\Delta ^2}{{\hat  X}_{s_{out}}(\phi_1)}} \rangle  = \langle {{\Delta ^2}{{ \hat X}_{i_{out}}(\phi_2)}} \rangle =1$, which means that the output noise level of SUI is the same as the vacuum state or coherent state even after the amplification of two OPAs.
Comparing with the conventional OPA scheme (see Eqs. (\ref{noise-amp})-(\ref{SNR-amp})), one sees that although the modulation signals of phase and amplitude, carried by probe beam, experience the same gain of $g=g_2=g_1$ in both cases, the noise of SUI is reduced by a factor of $1/(G_{1}^2+g_{1}^2)$ because of a destructive quantum interference effect for noise cancelation \cite{kong13}.

\begin{figure*}[htb]
\centering
 \includegraphics[width=16cm]{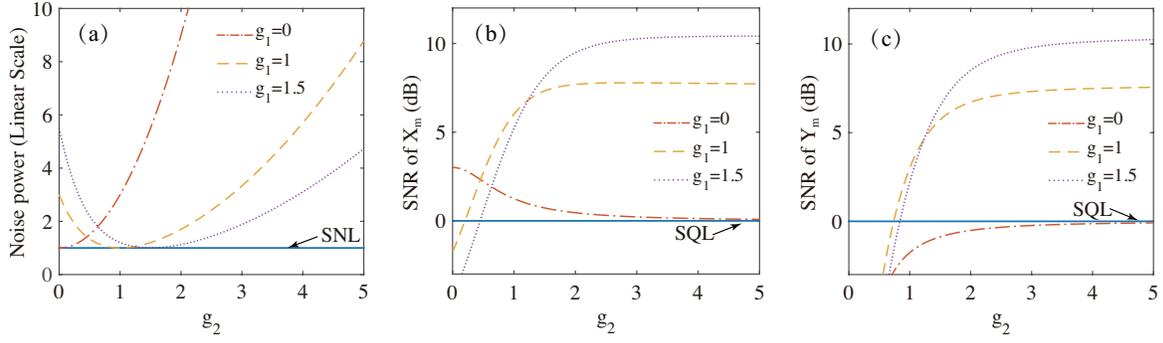}
\caption{
Noise power and SNRs of joint measurement as a function of $g_2$ when gain of OPA1 is fixed at different levels. 
(a) is the noise power at each output port of SUI, (b) and (c) are the SNRs of amplitude modulation signal $X_m=\epsilon$ and phase signal $Y_m=\delta$ respectively measured at the signal and idler outputs of SUI.
In the calculation, $I_{ps}\epsilon^2=1/2$; SNL, shot noise level; SQL, standard quantum limit.}\label{SNR-g2}

\end{figure*}

According to Eqs. (\ref{sig}) and (\ref{noise}), we have the SNRs of phase and amplitude simultaneously measured at the signal and idler outputs of the SUI as
\begin{equation}
\begin{split}
SNR_{SUI}(\hat X_s) = \frac{{4G_2^2{I_{ps}}\epsilon^2}}{{{{\left( {{G_2}{G_1} - {g_1}{g_2}} \right)}^2} + {{({G_1}{g_2} - {G_2}{g_1})}^2}}}\\
SNR_{SUI}(\hat Y_i) = \frac{{4g_2^2{I_{ps}} \delta^2}}{{{{\left( {{G_2}{G_1} - {g_1}{g_2}} \right)}^2} + {{({G_1}{g_2} - {G_2}{g_1})}^2}}},
\label{SNR}
\end{split}
\end{equation}
where the subscript ``SUI'' represents the quantum measurement scheme in Fig. \ref{fig-SUI}. For OPA1 with a fixed gain $g_1$, the maximum SNRs in signal and idler outputs are obtained at $g_2=2g_1G_1$ and $g_2\to \infty$, respectively. When the gain of OPA2 approaches infinity, i.e., ${g_2} \to \infty$, Eq. (\ref{SNR}) is rewritten as
\begin{equation}
\begin{split}
SNR_{SUI}(\hat X_s) = 2(G_1+g_1)^2{I_{ps}}\epsilon^2\\
SNR_{SUI}(\hat Y_i) = 2(G_1+g_1)^2{I_{ps}}\delta^2.
\label{SNR-m}
\end{split}
\end{equation}
Comparing Eq. (\ref{SNR-m}) with SQL (see Eq. (\ref{SNRopa})), it is obvious that for the probe field with fixed intensity $I_{ps}$, the SUI can achieve a better SNR than the classical OPA scheme with an enhancement factor of
\begin{equation}
\label{SNR improve}
\frac{SNR_{SUI}({\hat X_s})}{SNR_{Amp}({\hat X_s})} = \frac{SNR_{SUI}({\hat Y_i})}{SNR_{Amp}({\hat Y_i})} = (G_1+g_1)^2.
\end{equation}
This improvement factor in Eq. (\ref{SNR improve}) is originated from the noise cancelation due to the quantum correlations of the fields out of OPA1.

In addition to the joint measurement of two conjugate variables, such as phase and amplitude, SUI can simultaneously measure two quadrature amplitudes at arbitrary angles, i.e., $X_m(\theta) = X_m\cos\theta+ Y_m\sin\theta$ at $\theta = \theta_1,\theta_2$, with SNRs surpass the SQL when the LOs of HD1 and HD2 are properly adjusted. With the change of LO phase of $\phi_i$ ($i=1,2)$, one may expect a different, likely higher, noise level. This is true for the quantum scheme in Fig. \ref{fig-Qcode}, but is not the case for SUI. According to Eq. (\ref{noise}), the noise at the signal and idler output ports of SUI is irrelevant to the angle of quadrature amplitude. For the probe beam encoded with two quadrature-phase amplitudes $X_m(\theta_1)$ and $X_m(\theta_2)$, the homodyne measurement of $\hat X_s(\phi_1)$ and $\hat X_i(\phi_2)$ ($\phi_{1,2}=\theta_{1,2}$) at the two outputs will simultaneously decode the information of $X_m(\theta_1)$ and $X_m(\theta_2)$ with SNRs expressed as
\begin{equation}
\begin{split}
SNR_{SUI}(\hat X_s(\phi_1)) = 2(G_1+g_1)^2{I_{ps}}X_m(\theta_{1})^2\\
SNR_{SUI}(\hat X_i(\phi_2)) = 2(G_1+g_1)^2{I_{ps}}X_m(\theta_{2})^2.
\label{SNR_phi}
\end{split}
\end{equation}
Comparing with the SQL, it is clear that the improvement factor of SNRs in Eq. (\ref{SNR_phi}) is the same as in Eq. (\ref{SNR improve}).

It is worth noting that for the SUI scheme, the optimum condition for achieving the absolute minimum noise at the two output ports is different from that for obtaining the maximum SNRs in joint measurement. For OPA1 with a fixed gain $g_1$, the former is achieved for OPA2 with the same gain as OPA1 ($g_1=g_2$), while the latter is obtained for OPA2 with gain approaching infinity (${g_2} \to \infty$).
To better understand this difference, we calculate from Eqs. (\ref{noise}) and (\ref{SNR}) the noise and SNRs measured at each output of SUI when $g_1$ takes different values. The results in Fig. \ref{SNR-g2} is calculated by assuming $I_{ps}\epsilon^2=1/2$.
Since the noise powers at signal and idler outputs are same (see Eq. (\ref{noise})), so we calculated the noise at signal output and show the results in Figs. \ref{SNR-g2}(a). 
Moreover, we plot the SNR as a function of $g_2$ to show the variation trend of simultaneously measured of amplitude and phase modulation, $X_m, Y_m$ in Figs. \ref{SNR-g2}(b) and \ref{SNR-g2}(c), respectively. 
For the ease of comparison, the corresponding SNL ($\langle {{\Delta ^2}{{\hat X}(\phi)}} \rangle=1 $) and SQL of SNRs obtained by substituting $g_1=0$ and ${g_2} \to \infty$ in Eqs. (\ref{noise}) and (\ref{SNR}) are depicted in Fig. \ref{SNR-g2} as well.
It is clear that minimum noise at SNL is achieved for $g_2 = g_1$, at which the SNRs also beat SQL. 
However, the highest SNRs for a given $g_1$ is not obtained under the condition of $g_1=g_2$. For a fixed $g_1$ ($g_1\neq 0$), the SNR continues to increase with $g_2$ when $g_2 > g_1$, and the optimum SNR is obtained under the condition of ${g_2} \gg g_1$. 
From Fig. \ref{SNR-g2}, we find that the optimum SNR depends only on $g_1$ and is  better than the SNR at $g_1=g_2$ by about 3 dB when $g_1\gg 1$.

Notice that the optimum improvement factor of SNR for SUI given in Eq. (\ref{SNR improve}) is the same as that for the quantum dense coding scheme in Eq. (\ref{SNR-EPR1}). 
Why the optimum SNR is obtained at $g_2\rightarrow \infty$? We think this is because OPA2 functions as a 50/50 beam splitter  when $g_2\rightarrow \infty$. 
Therefore, in the sense of coherently mixing the two correlated fields, the role of OPA2 operated in the high gain regime is the same as the 50/50 BS in Fig. \ref{fig-Qcode}.

In the discussion above, we set the operating point of the SU(1,1) interferometer at the dark fringe by adjusting the overall phase of the interferometer to $\varphi=\pi$. 
This is because the maximum quantum noise reduction occurs at dark fringe due to destructive quantum interference. Thus, $\varphi=\pi$ should be the optimum operating point for the SUI. Since we just showed that the SUI with $g_2 \gg g_1$ is equivalent to the quantum dense coding scheme in Fig. \ref{fig-Qcode}, this optimum operating point can also be viewed as an equivalent of the optimum squeezing at the two outputs of the BS in Fig. \ref{fig-Qcode}.

\begin{figure}[htb]
\centering
 \includegraphics[width=6.5cm]{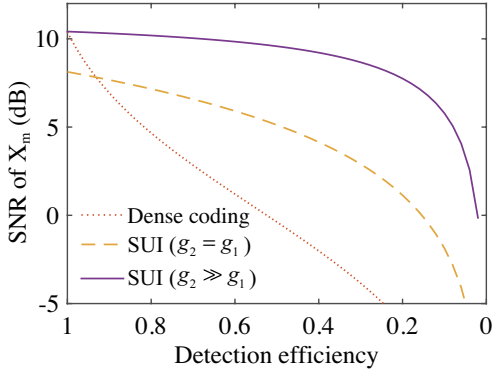}
\caption{The SNR of measured amplitude modulation of $X_m$ versus detection efficiency for both SUI and dense coding scheme.
In the calculation, $I_{ps}\epsilon^2=1/2$; SUI is working under the condition of $g_1=g_2=1.5$ (dashed curve) and $g_2=5 \gg g_1=1.5$ (solid curve), respectively; the gain of OPA in dense coding scheme is $g=1.5$; and the corresponding SQL of joint measurement is 0 dB.
}\label{SUI-SNR-loss}
\end{figure}

The improvement factors of the quantum enhanced joint measurement for the two schemes in Figs. \ref{fig-Qcode} and \ref{fig-SUI} are the same, but SUI surpasses the dense coding scheme in three aspects. 
Firstly, SUI can utilize the EPR correlation between two fields with different wavelengths. 
Secondly, the noise reduction of SUI does not depend on the phase of LO in homodyne detection and the improvement in SNRs does not vary with the angles of the quadrature-phase amplitudes encoded on probe. 
Thirdly, the influence of the detection loss on SNR is diminished because the vacuum noise introduced through loss is negligible compared to the noise at the outputs of SUI (see Fig. \ref{SNR-g2}(a) and Eq. (\ref{noiseloss})) \cite{PDL12}. 
The first two points have been well presented in the analysis above. 
To illustrate the third advantage, we plot the SNRs as a function of detection efficiency when the gain of OPA2 in SUI is respectively set to achieve optimum SNR ($g_2 \gg g_1$, solid curve) and lowest noise ($g_1=g_2$, dashed curve) by taking the jointly measured amplitude as an example, as shown in Fig. \ref{SUI-SNR-loss}. 
In the calculation, similar to the deduction of Eqs. (\ref{Sigloss})-(\ref{noiseloss}), the non-ideal detection is modeled as an insertion loss of BS placed in front of HD. As a comparison, the relation between SNR and detection efficiency for the dense coding scheme is depicted in Fig. \ref{SUI-SNR-loss} as well. 
One sees that with the decrease of detection efficiency, the value of SNR for the dense coding scheme quickly decreases (dotted curve), while the downtrend of SNRs for SUI scheme is very slow, particularly for the SUI with $g_2 \gg g_1$. 
Previous experimental demonstration of the loss insensitive property was performed by the SUI under the operation condition of $g_2 = g_1$ \cite{hud14,chek17}. Our results in Fig. \ref{SUI-SNR-loss} indicate that comparing SUI with $g_1=g_2$, the SUI with $g_2 \gg g_1$ is not only able to achieve another 3 dB improvement, but posses a better loss-tolerance feature.

This loss-insensitive property can be used to split the signal and idler outputs further into more beams for the joint measurement of multiple quantities without a significant reduction of the SNRs.
Therefore, SUI can directly accomplish the quantum enhanced joint measurement for arbitrary number of non-commuting observables. 
For example, if we further split the signal output into two with a 50/50 BS (see the dashed box in Fig. \ref{fig-SUI}) and place another set of HD (HD3) at the reflection port, we can realize the joint measurement of three non-commuting quadrature-phase amplitudes $ X_m(\theta_1)$, $ X_m(\theta_2)$, and $ X_m(\theta_3)$ with sensitivity beyond SQL by using HD1, HD2 and HD3 to simultaneously perform measurement. 
For example, if we have $g_1=1$ and $g_2=5$ and each HD device is perfect, the calculated SNRs of mesured $X_m(\theta_1)$, $ X_m(\theta_2)$, and $ X_m(\theta_3)$ surpass the SQL by 7.25 dB, 7.6 dB and 7.25 dB, respectively. The reduction of SNRs in the ports split by BS is only 0.35 dB lower than that in the port without splitting.

\subsection{Post-detection processing for the joint measurement of multiple parameters}

The ability of SUI scheme to make a measurement of a modulation signal at arbitrary angle $X_m(\theta) = X_m \cos \theta+ Y_m \sin \theta$ can be achieved indirectly through the method of post-detection processing as well. 
The basic principle of the method is to measure a pair of conjugated quadrature amplitudes $X_m$ and $Y_m$ by using HD1 and HD2 at signal and idler output ports.
The measurement of $X_m(\theta)=X_m \cos \theta +Y_m \sin \theta$ with a modulation depth $\gamma=X_m(\theta)$ is then achieved by processing the  photocurrents out of the two sets of HDs.
The information $X_m(\theta)$ encoded on the probe can be decomposed into phase and amplitude modulations at the same frequency. 
So the complex amplitude of the probe field is proportional to $1 + i \delta -\epsilon$ with $\delta  = \gamma \sin \theta$ and $\epsilon  = \gamma \cos \theta$. When the two orthogonal quadratures $X_m$ and $Y_m$ are obtained by measuring $\hat X_s$ and $\hat Y_i$ with HD1 and HD2 at the two outputs of SUI, we have
\begin{eqnarray}
\hat X_{s} &=& \hat a_{s_{out}}e^{-j\varphi_0} + \hat a_{s_{out}}^\dag e^{j\varphi_0}\cr &=& {G_2}{\hat X_{s_{out1}}} - {g_2}{\hat X_{i_{out1}}} - {G_2}{\hat X_{s_{out1}}}\epsilon  - {G_2}{\hat Y_{s_{out1}}}\delta \cr
\hat Y_{i} &=& (\hat a_{i_{out}}e^{-j\varphi_0} - \hat a_{i_{out}}^\dag e^{j\varphi_0})/j \cr &= & - {G_2}{\hat Y_{i_{out1}}} - {g_2}{\hat Y_{s_{out1}}} - {g_2}{\hat X_{s_{out1}}}\delta  + {g_2}{\hat Y_{s_{out1}}}\epsilon , \cr &&
\end{eqnarray}
where $e^{j\varphi_0}\equiv \alpha/|\alpha|$ is the phase of the seed injection.
Using the relation $\hat X_{\theta} \equiv \cos \theta \hat X_{s} + k\sin \theta \hat Y_{i}$, where $k$ is a coefficient that balance the gain difference between the signal and idler ports, we obtain the average signal power and noise fluctuation for the measurement of signal $X_m(\theta)$:
\begin{eqnarray}
\label{Signal_arb_signal}
\langle \hat X_{\theta} \rangle ^2 &=& \left\langle {(\cos \theta {\hat X_{s}} + k\sin \theta {\hat Y_{i}})} \right\rangle ^2  \cr &=&4(G_2 \cos ^2\theta  + k g_2 \sin ^2\theta)^2 I_{ps}{\gamma ^2}
\end{eqnarray}
and
\begin{eqnarray}
\label{Noise_arb_signal}
\langle \Delta \hat X^2_{\theta} \rangle
&=& \left[({G_2}{G_1} - {g_1}{g_2} )^2 + ({G_1}{g_2} - {G_2}{g_1})^2\right]\cr
&&\hskip 0.5in \times(\cos^2\theta  + {k^2}\sin^2\theta ).
\end{eqnarray}
Consequently, we have the SNR
\begin{eqnarray}
\label{SNR-post}
&& SNR(\hat X_{\theta})=\cr && \frac{4(G_2 \cos ^2\theta  + k g_2 \sin ^2\theta)^2 I_{ps}{\gamma ^2}}{\left[({G_2}{G_1} - {g_1}{g_2} )^2 + ({G_1}{g_2} - {G_2}{g_1})^2\right](\cos^2\theta  + {k^2}\sin^2\theta )}\cr &&
\end{eqnarray}
In the case of $k=G_2/g_2$ and $g_2\to \infty$, Eq. (\ref{SNR-post}) has the simplified form
\begin{equation}
\label{SNR-post-s}
SNR(\hat X_{\theta}) = 2{\left( {{G_1} + {g_1}} \right)^2}{I_{ps}}{\gamma ^2},
\end{equation}
which indicates that quantum enhanced factor obtained by the post-detection is the same as that by using direct detection (see Eq. (\ref{SNR_phi})).

The method of post-detection data processing can be extended for the joint measurement of multiple modulation signals. 
For example, if the probe beam carries the modulation information in $N$ different quadrature amplitudes, the measurement of $ X_m(\theta_1)$, $ X_m(\theta_2)$, $ X_m(\theta_3)$ $\cdots$, and $ X_m(\theta_N)$ can be simultaneously obtained from the calculation of the photocurrent $i(\theta) = i_{1} \cos\theta + k i_{2}\sin\theta$ after substituting $\theta$ with $\theta_1$, $\theta_2$, $\cdots$, $\theta_N$, where $i_{1}$ and $i_{2}$ are the photocurrents out of the HD1 and HD2 when the measurements of $\hat X_{s}$ and $\hat Y_{i}$ are simultaneously performed. 
Comparing with the method of directly detecting each quadrature amplitude with a HD, which we have discussed in the end of Sec. IIIB, the method of post-detection data processing seems more convenience for realizing the joint measurement of non-commuting observables with number greater than $3$ because there is no need to increase the number of HD devices. 
However, in practice, the influence of detection efficiency and the noise correlation between the two outputs of SUI may introduce extra complexity in the post-detection processing.

\subsection{SU(1,1) Interferometer with Dual-beam Sensing}

\begin{figure}[htb]
\centering
 \includegraphics[width=8cm]{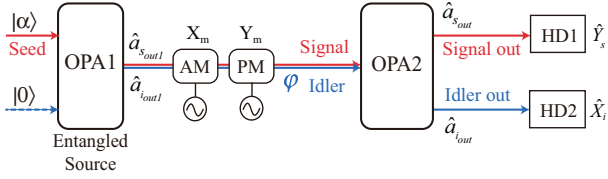}
\caption{SU(1,1) interferometer with dual beams function as the sensing field. OPA, optical parametric amplifier; AM, amplitude modulator; PM, phase modulator; HD, homodyne detection.  }\label{dual}
\end{figure}

One unique property of SU(1,1) interferometer is that the interference fringe depends on the sum of the phases of the signal and idler beams between two OPAs \cite{hud14,Guo2016}. This suggests that passing both signal and idler fields out of OPA1 through the modulation units of AM and PM will double the signal size at the outputs of SUI. This idea is shown in Fig. \ref{dual}. Different from the SUI in Fig. \ref{fig-SUI}, in which the signal field out of OPA1 functions as the sensing field, the sensing field in Fig. \ref{dual} is the two non-degenerate fields produced by OPA1.

In Fig. \ref{dual}, when the information of phase and amplitude ($Y_m=\delta$ and $X_m=\epsilon$) is encoded on the dual beams out of OPA1 and OPA2 is operated in the deamplification condition, we analyze the SNRs of $Y_m$ and $X_m$ by using HDs at signal and idler outputs to respectively measure $\hat Y_s$ and $\hat X_i$. We first deduce the intensity of phase signal at the signal output port
\begin{eqnarray}
\label{phase-dual}
\langle \hat Y_s\rangle^2 = 4(G_1G_2+g_1g_2)^2 |\alpha|^2\delta^2.
\end{eqnarray}
Since the noise level at the outputs of SUI in Fig. \ref{dual} is the same as that in Fig. \ref{fig-SUI}, i.e., $\langle \Delta^2\hat Y_s\rangle = (G_2G_1-g_2g_1)^2+(G_2g_1-g_2G_1)^2$, the SNR of $Y_m$ measured at the signal output port is
\begin{eqnarray}
\label{phase-dual-SNR}
&& SNR_{DB}(\hat Y_s) = \cr && \hskip 0.3 in {4(G_1G_2+g_1g_2)^2 I_{ps} \delta^2\over (G_1^2+g_1^2)[(G_2G_1-g_2g_1)^2+(G_2g_1-g_2G_1)^2]}\cr
&& \hskip 0.3 in \rightarrow 4(G_1+g_1)^2I_{ps}\delta^2 ~~{\rm for}~~ g_1 \gg 1~ {\rm and}~ g_2\gg g_1,~~
\end{eqnarray}
where the subscript ``DB'' refers to the dual-beam scheme, and the photon number of sensing field $I_{ps} = (G_1^2+g_1^2)|\alpha|^2~ (|\alpha|^2\gg1)$ is the total intensity of the signal and idler fields out of OPA1. Moreover, under the same condition, it is straight forward to calculate the intensity of amplitude signal at the idler output port
\begin{eqnarray}
\label{amplitude-dual-SNR}
\langle\hat X_i\rangle^2 =  {4(G_1g_2-g_1G_2)^2 I_{ps} \epsilon^2},
\end{eqnarray}
which shows that the intensity of amplitude modulation at idler output of SUI is negligibly small. Particularly, for the case of $g_1\gg1, g_2\gg1$, we have $\langle\hat X_i\rangle^2 \to 0$. 
This is caused by the common mode rejection in intensity fluctuation when OPA2 in Fig. \ref{dual} is operated at deamplification condition. 
Indeed, each output field of SUI in Fig. \ref{dual} carries the same information. 
If we measure $\hat Y_i$ in the idler output, the same SNR as measuring $\hat Y_s$ for the phase modulation can be obtained. So $\hat Y_i$ can be viewed as an exact copy of $\hat Y_s$, which means the phase information carried by the sensing field can be split into two without adding noise \cite{Guo2016}.

The above results indicate that the dual-beam scheme in Fig. \ref{dual} cannot realize the quantum enhanced joint measurement of phase and amplitude modulations. However, for the phase measurement only, the improvement factor over SQL is twice that of SUI in Fig. \ref{fig-SUI}.

\subsection{A Rule of Resource Conservation for Joint Measurement}

Comparing the measurement realized by using the quantum schemes in Fig. \ref{dual} and Fig. \ref{fig-SUI}, we find the relation $SNR_{SUI}(\hat X_s) + SNR_{SUI}(\hat Y_i) = SNR_{DB}(\hat Y_s)$ holds for the case of equal modulation strength $\delta=\epsilon$, which means that the total sum of the SNRs for joint measurement equals the SNR for one observable obtained by consuming all the resource on it. In fact, the comparison of the classical measurement schemes in Figs. \ref{fig-DM}-\ref{fig-Amp} also reveals this resource partition relation, as shown in Eqs. (\ref{SNR-X+Y}) and (\ref{SNR-res}).

\begin{figure}[htb]
\centering
 \includegraphics[width=8cm]{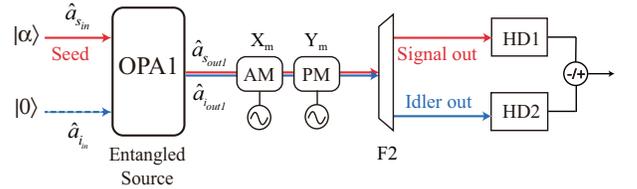}
\caption{Measurement scheme of using entanglement to measure one observable as precise as possible. AM, amplitude modulator; PM, phase modulator; OPA, optical parametric amplifier; F2, dualband filter; HD, homodyne detection. }\label{DM-EPR}
\end{figure}

To better understand the resource conservation rule discussed above, let us consider if the rule applies for the quantum resource consumed in the dense coding scheme in Fig. \ref{fig-Qcode}. Hence, we try to figure out a new scheme in which the entanglement generated by an OPA is fully used to measure one observable---phase or amplitude---as precise as possible. As shown in Fig. \ref{DM-EPR}, the entangled signal and idler fields co-propagate and function as the probe for carrying information. After sending the dual-beam, encoded with the phase and amplitude signals ($Y_m=\delta$ and $X_m=\epsilon$) by weakly modulating PM and AM, through a dualband filter F2, signal and idler fields are then separated and respectively detected by HD1 and HD2. The quadrature amplitudes $\hat X_{s,i}$ and $\hat Y_{s,i}$ of signal and idler fields are measured by setting the two HDs at $\phi_{1,2}=0$ and $\phi_{1,2}=\pi/2$, respectively. To reduce the noise level in the measurement, we combine the photocurrents of two HDs with subtractor or adder so that the quantum noise cancelation due to quantum correlation via $\hat X_s - \hat X_i$ and $\hat Y_s+\hat Y_i$ can be utilized.

When the photocurrents of HD1 and HD2 with phase of LOs locked at $\phi_{1,2}=\pi/2$ are added, we have the power of the phase signal $Y_m$
\begin{equation}
\label{Sig-dbdm}
\langle \hat Y_s + \hat Y_i \rangle ^2 = 4 (G+g)^2 |\alpha|^2 \delta^2,
\end{equation}
The corresponding noise is
\begin{equation}
\label{noise-dbdm}
\langle \Delta (\hat Y_s + \hat Y_i) \rangle ^2 = 2 (G- g)^2 = \frac{2}{(G+ g)^2}
\end{equation}
From the Eqs. (\ref{Sig-dbdm}) and (\ref{noise-dbdm}), it is straightforward to obtain SNR for the measured phase modulation
\begin{eqnarray}
\label{DM-EPR-SNR}
SNR_{DB-DC}(\hat Y) = && \frac{2 (G+ g)^4 I_{ps} \delta^2}{G^2 + g^2}\cr
&& \hskip -0.3 in \rightarrow 4(G+g)^2I_{ps}\delta^2 ~~{\rm for}~~ g \gg 1,~~~~
\end{eqnarray}
where the subscript ``DB-DC'' represents the scheme in Fig. \ref{DM-EPR}.
On the other hand, when the photocurrents of HD1 and HD2 with phase of LOs locked at $\phi_{1,2}=0$ are subtracted, the measurement noise $\langle \Delta (\hat X_s -\hat X_i) \rangle ^2 = \frac{2}{(G+ g)^2}$ is lower than the shot noise level. However, in this case, the power of amplitude signal $X_m$, which can be express as $\langle \hat X_s - \hat X_i \rangle ^2 = 4 (G-g)^2 |\alpha|^2 \epsilon^2$, is negligibly small, particularly, in the high gain regime.
Therefore, the measurement results are the same as those in Eqs.(\ref{phase-dual-SNR}) and (\ref{amplitude-dual-SNR}) given in Sec. IIID.

Notice that the scheme in Fig. \ref{DM-EPR} can only be used to measure the phase with sensitivity beyond SQL. Comparing with dense coding scheme for joint measurement in Fig. \ref{fig-Qcode}, we find the relation $SNR_{DC}(\hat X_{b_1})+SNR_{DC}(\hat Y_{b_2}) = SNR_{DB-DC}(\hat Y)$ holds for the case of $\delta=\epsilon$, which again demonstrates the joint measurement can be viewed as the partition of total resources, i.e., the fixed amount of noise reduction factor of $1/(G+g)^2$ originated the EPR-correlated source (OPA1) is distributed among the measurement of two jointly measured quantities.

\section{Heisenberg Limit for the joint measurement of phase and amplitude}

It is known from the very beginning that the sensitivity of SU(1,1) interferometer in phase measurement is bounded by the Heisenberg limit  when there is no intra-interferometer loss and no seed injection \cite{yur,PDL17b,ou12}. Now we will study the Heisenberg limit for the joint measurement realized by using SU(1,1) interferometer in Fig. \ref{fig-SUI}.

In order to reach Heisenberg limit in the joint measurement, similar to the phase measurement \cite{yur,ou12}, the OPA1 of SUI must be operated without input, i.e., the intensity of the seed input field in Fig. \ref{fig-SUI} is $\left| \alpha  \right|^2=0$. In this case, the photon number of probe signal beam is $N=I_{ps}=g_1^2$, and the measured powers of amplitude and phase signals in Eq. (\ref{sig}) are rewritten as
\begin{equation}
\langle \hat X_{s_{out}}\rangle^2  = 4G_2^2{g_1^2}\epsilon^2,~~~~ \langle \hat Y_{i_{out}}\rangle^2  =4g_2^2{g_1^2}\delta^2.
\label{sig-H}
\end{equation}
In this case, for the SUI operated at the deamplification condition with ${g_2} \to \infty$, the SNRs of measured phase and amplitude in Eq. (\ref{SNR}) have the simplified form:
\begin{equation}
\begin{split}
SNR_{\delta} = {\langle \hat X_{s_{out}}\rangle^2\over \langle \Delta^2 \hat X_{s_{out}}^2\rangle}   =2{\left( {{G_1} + {g_1}} \right)^2}g_1^2{\delta ^2}\\
SNR_{\epsilon} = {\langle \hat Y_{i_{out}}\rangle^2\over \langle \Delta^2 \hat Y_{i_{out}}^2\rangle}   =2{\left( {{G_1} + {g_1}} \right)^2}g_1^2{\epsilon ^2}.
\label{SNR-H-limit}
\end{split}
\end{equation}
Under the condition of $SNR_{\delta}=SNR_{\epsilon}=1$ and $g_1 \gg 1$, we obtain the minimum detectable amplitude and phase signals encoded on the probe signal beam, $\epsilon_m = 1/2\sqrt{2}N$ and $\delta_m = 1/2\sqrt{2}N$, which are the Heisenberg limit of the joint measurement.

\begin{figure}[htb]
\centering
 \includegraphics[width=6.5cm]{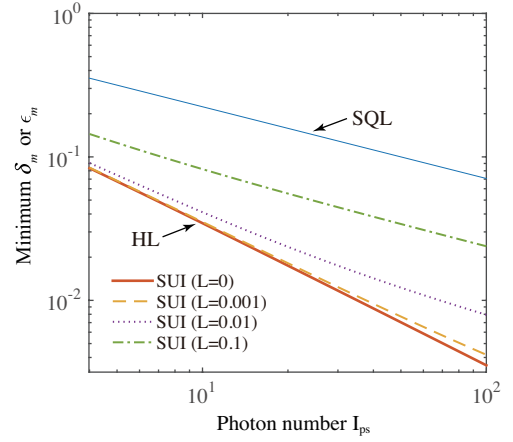}
\caption{Minimum measurable amplitude and phase signals, $\epsilon_m$ and $\delta_m$, as a function of photon number of the probe field $I_{ps}$ for SU(1,1) interferometer with different internal losses $L$. SQL, standard quantum limit; HL, Heisenberg limit. }\label{loss}
\end{figure}

On the other hand, when the losses inside the interferometer (transmission losses between two OPAs) are included, the performance of SUI will be severely affected. For simplicity, assuming the losses of the two arms in SUI, labeled as signal and idler, are equal, we then model the loss $L$ as a beam splitter with transmissivity of $T=1-L$. After some algebra, it is straight forward to deduce powers of the amplitude and phase modulation measured at the signal and idler outputs:
\begin{eqnarray}
\langle \hat X_s\rangle^2 = 4(1-L)G_2^2g_1^2\epsilon^2, ~~\langle \hat Y_i\rangle^2 = 4(1-L)g_2^2g_1^2\delta^2,~~~~
\end{eqnarray}
and the corresponding noise fluctuations are given by
\begin{eqnarray}
&&\langle \Delta^2\hat X_s\rangle = \langle \Delta^2\hat Y_i\rangle\cr
&& \hskip 0.3 in = (1-L) \left[ ( {{G_2}{G_1} - {g_1}{g_2}} )^2 + ({G_1}{g_2} - {G_2}{g_1})^2\right] \cr && \hskip 0.8 in + L(G_2^2 +g_2^2).
\end{eqnarray}
Under the condition of $g_2\rightarrow \infty$, the SNRs of measured amplitude and phase are then expressed as
\begin{equation}
\begin{split}
SNR_{SUI-Loss}(\hat X_s) = \frac{2 (1-L)I_{ps} \epsilon^2}{(1-L)(G_1 - g_1)^2 + L}\\
SNR_{SUI-Loss}(\hat Y_i) = \frac{2 (1-L)I_{ps} \delta^2}{(1-L)(G_1 - g_1)^2 + L}
\end{split}
\end{equation}
where $I_{ps} = g_1^2$. Setting the values of SNRs to1, we find the minimum measurable modulation of amplitude and phase signals are
\begin{eqnarray}
\epsilon_m = \delta_m &=& \sqrt{ \frac{1}{2 I_{ps}(G_1+g_1)^2} + \frac{L}{2(1-L)I_{ps}}}\cr
&=& \sqrt{ \frac{1}{2 I_{ps}^2(1+\lambda)^2} + \frac{L}{2(1-L)I_{ps}}}
\label{minimum-loss}
\end{eqnarray}
with $\lambda\equiv G_1/g_1 = \sqrt{(I_{ps}+1)/I_{ps}}$. To demonstrate the influence of loss $L$, we plot the minimum measurable signal, $\epsilon_m$ or $\delta_m$, as a function of $I_{ps}$ when the value of $L$ is different, as shown in Fig. \ref{loss}. For the convenience of comparison, we also depict the SQL ($1/\sqrt{2 I_{ps}}$) and HL (achieved by substituting $L=0$ in Eq. (\ref{minimum-loss})) in Fig. \ref{loss}.
It is obvious that for both the classical limit of SQL and quantum limit of HL, the sensitivity, represented by the minimal measurable $\epsilon_m$ and $\delta_m$, increase with the photon number of probe field $I_{ps}$. 
When the intensities of sensing fields are the same, the ratio of the sensitivity for HL and SQL increase with $I_{ps}$. 
In other word, the quantum enhancement factor increases with $I_{ps}$.
Moreover, we find that for the SUI with internal loss $L$, the Heisenberg limit of $\sim 1/2\sqrt{2}I_{ps}$ is achievable only when $I_{ps}< 1/L$, whereas the SQL of $\sim 1/\sqrt{2 I_{ps}}$ will be approached when $I_{ps} \gg 1/L$. 
The results indicate that if the photon number of sensing field $I_{ps}$ is low, the deduction of the quantum enhanced factor is not significant. 
However, if the photon number of $I_{ps}$ is high, the quantum enhancement factor will be dramatically affected by the loss $L$. For example, when the photon number of $I_{ps}$ is 4, $10\%$ loss inside the SUI only increase $\epsilon_m$ and $\delta_m$ by about 1.7 times that of HL. 
On the other hand, when the photon number of $I_{ps}$ is 100, $10\%$ loss will increase $\epsilon_m$ and $\delta_m$ by about 6.8 times that of HL, although the increase of  $I_{ps}$ from 4 photon to 100 photons increase the sensitivity by about 6 times. 
Therefore, for the practical application, the SUI with a stronger sensing field is preferred but its sensitivity for joint measurement is limited by the losses inside the interferometer. This is consistent with previous studies \cite{ou12,dkg12}.

\section{Conclusion}

In summary, we have investigated various schemes for joint measurement of multiple non-commuting observables with both classical sources and quantum sources and compared their performance under the condition of the same probe intensity. 
We find that quantum schemes using entangled sources have significant improvement in signal-to-noise ratio over the classical schemes. 
The dense coding scheme with frequency degenerated entanglement as the quantum source is vulnerable to losses. However, the newly developed SU(1,1) interferometer having the ability of coherently mixing two entangled fields with different wavelengths, is insensitive to the losses outside the interferometer such as propagation and detection losses. Moreover, the SU(1,1) interferometer can be extended for joint measurement of multiple ($>2$) quadrature-phase amplitudes with arbitrary angles. Furthermore, in our investigation, we find an interesting effect of resource partition in the joint measurement of two orthogonal observables, that is, the sum of the SNRs for joint measurement of two orthogonal observables is equal to the SNR of one observable measured by consuming all the measurement resource on it. This rule applies to both classical and quantum schemes.

\section*{ACKNOWLEDGMENTS}
The work is supported in part by the National Key Research and Development Program of China (2016YFA0301403), 973 program of China (2014CB340103), National Natural Science Foundation of China (91736105, 11527808), and by the 111 project  B07014.

\end{document}